\documentclass[prl,twocolumn,showpacs,preprintnumbers,amsmath,amssymb]{revtex4}

\usepackage{multirow}
\usepackage{graphicx}
\usepackage{dcolumn}
\usepackage{bm}
\newcommand{\lco}{La$_2$CuO$_4$}

\newcommand{\cbo}{CuB$_2$O$_4$}

\begin{document}

\title{Aspects of resonant inelastic X-ray scattering in quasi-zero-dimensional CuB$_2$O$_4$}

\author{J. N. Hancock$^{1}$}
\author{G. Chabot-Couture$^{2}$}
\author{Y. Li$^{3}$}
\author{G. A. Petrakovski\u{\i}$^{4}$}
\author{K. Ishii$^{5}$}
\author{I. Jarrige$^{5}$}
\author{J.-i. Mizuki$^{5}$}
\author{T. P. Devereaux$^{1}$}
\author{M. Greven$^{1,2}$}

\affiliation{$^{1}$Department of Photon Science and Stanford Synchrotron Radiation Laboratory, Stanford, California 94309}
\affiliation{$^{2}$Department of Applied Physics, Stanford, California 94305}
\affiliation{$^{3}$Department of Physics, Stanford, California 94305}
\affiliation{$^{4}$Kirenski\u{\i} Institute of Physics, Siberian Division, Russian Academy of Sciences, Krasnoyarsk, 660036 Russia}
\affiliation{$^{5}$Synchrotron Radiation Research Center, Japan Atomic Energy Agency, Hyogo 679-5148, Japan}

\date{\today}

\begin{abstract}
We present a resonant inelastic scattering (RIXS) study using \cbo, a lattice of CuO$_4$ plaquettes electronically isolated by B$^{+3}$ ions. The observed Cu $K$-edge spectra show a small number of well-separated features, and the simple electronic structure of \cbo\ allows us to explore RIXS phenomenology. We find a low energy feature that cannot be attributed to the same charge-transfer excitation discussed in other cuprates, and is likely a $d$$\rightarrow$$d$ transition thought to be forbidden under common considerations of $K$ edge RIXS. 
\end{abstract}

\pacs{PACS}

\maketitle

New experimental techniques are often born from the advent of new technology, as is the case of high-energy resolution ($\sim$100 meV) resonant inelastic X-ray scattering (RIXS), which has been enabled by the development of the diced analyzer crystal and other synchrotron advances. RIXS is unique in its atomic-species-specific and momentum-selective ability to probe electronic excitations in strongly correlated electron systems  \cite{hill98,abba99,hasan00,lu05,lu06,ghir04,dallera02, platzman98} and can, in principle, couple to both local \cite{abba99,hill98,platzman98} and extended \cite{hasan00,lu05,lu06} electronic excitations, including $d$$d$ excitations \cite{ghir04} and coupled magnetic excitations \cite{hill08}. Unfortunately, detailed quantitative analysis of experimental data has so far been limited in scope.

In this Communication, we present a comprehensive experimental and theoretical study of RIXS scattering at the Cu $K$ edge of the model quasi-zero-dimensional compound \cbo. Among our findings is a low-energy feature which cannot be attributed to the charge-transfer excitation widely discussed in higher-dimensional cuprates\cite{abba99,hasan00,lu05,lu06}, but more likely represents a $dd$ excitation. This anomaly would be nominally symmetry forbidden according to conventional theory. The identification of an unexpected RIXS scattering channel has potential implications to higher-dimensional compounds, such as the high-$T_c$ superconductors, and for the selection rules of indirect RIXS in general, opening avenues for the interpretation of existing data.

\cbo\ consists of unlinked CuO$_4$ subunits that are isolated electronically by B$^{+3}$ ions, while the Cu retain a nominal +2 valence, and serves as a `molecular' analog to the `CuO$_2$ planes' of the high-$T_c$ parent compounds. \cbo\ forms an interesting magnetic phase \cite{petra00,roessli01} and exhibits a remarkably strong magneto-optical effect \cite{saito08}, but these phenomena occur below 25 K and involve very low-energy (magnetic) degrees of freedom.

\begin{figure}
\begin{center}
\includegraphics[width=3.3in]{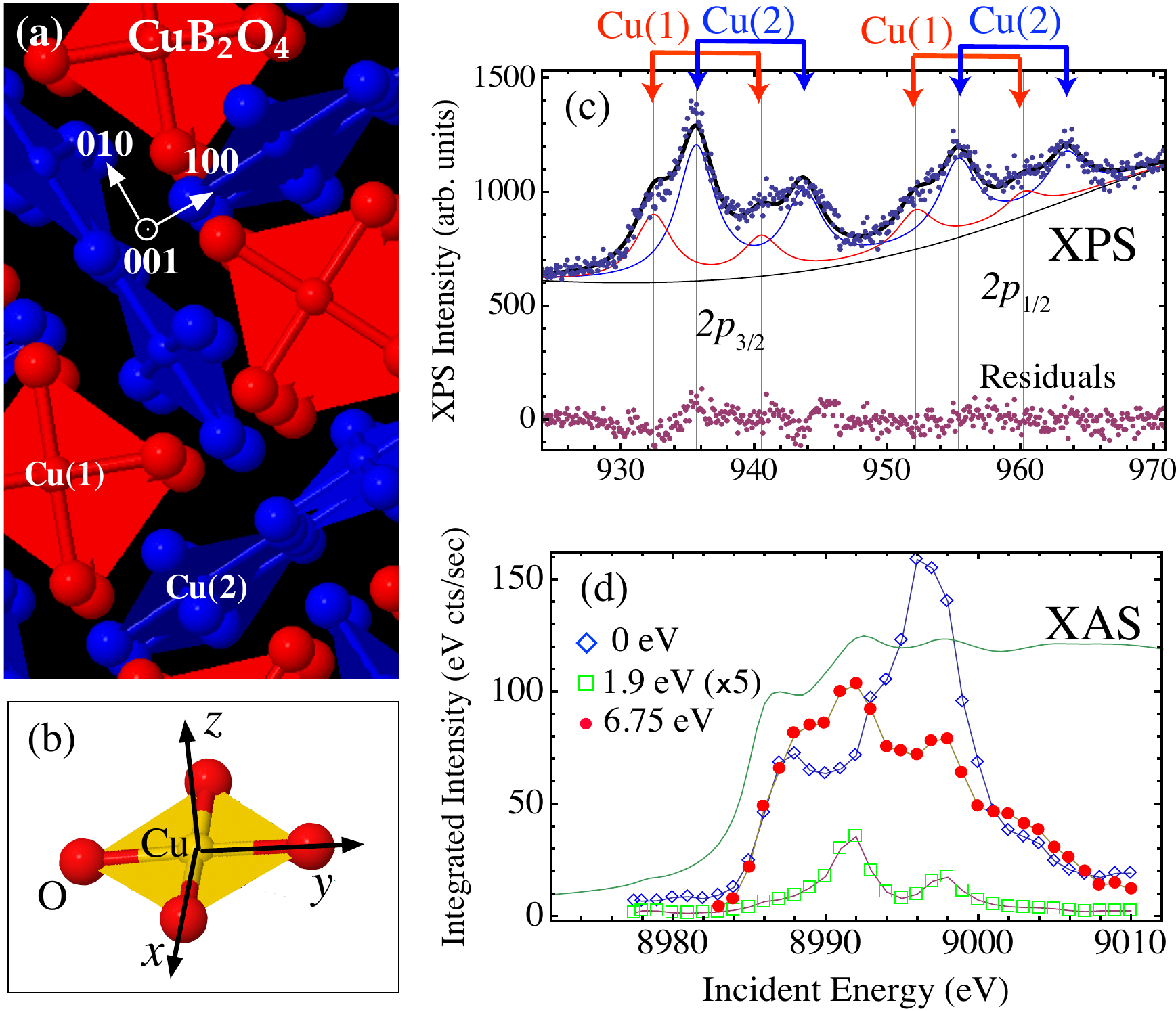}
\caption{(Color) (a) Schematic structure of \cbo, with undistorted Cu(1) plaquettes (red) and distorted Cu(2) plaquettes (blue). B atoms are not shown. (b) Elementary CuO$_4$ unit and local coordinate system. (c) XPS spectrum. Lines are the result of a fit to Lorentzian lineshapes and a constant slope plus broad Lorentzian to represent the background. (d) XAS (green line) and integral of corrected RIXS intensity (symbols) versus incident photon energy.}
\label{fig1}
\end{center}
\end{figure}

Figure \ref{fig1}a shows a $\langle 001 \rangle$ view of part of the crystal structure of CuB$_2$O$_4$, which contains two crystallographically inequivalent CuO$_4$ plaquettes. Figure \ref{fig1}c shows Cu $2p_\frac{3}{2}$ and $2p_\frac{1}{2}$ X-ray photoemission spectroscopy (XPS) spectra collected on a sample cleaved in-situ under 10$^{-8}$ torr. Each of the four main features is composed of two peaks, with the stronger peak at higher binding energy, and an intensity ratio of about 1:2. The data are well described assuming two sets
of four peaks arising from the Cu(1) and Cu(2) sites, respectively, with a shift of
$\sim$3.2 eV which could reflect a chemical shift between these sites. The two peaks within a multiplet (e.g., 2$p_{\frac{3}{2}}$) and for one particular site (e.g., Cu(1)) represent the overlap of core hole states with the ground state \cite{vanv93}. The stronger peak at lower binding energy corresponds to the ``well-screened" (WS) state with mostly $d^{10}\underline{L}$ character, while the other peak corresponds to the ``poorly screened" (PS) state with predominant $d^{9}$ character. An identification of related core hole states is possible in the higher-dimensional systems, where it is known that nonlocal (inter-plaquette) screening channels play a role \cite{vanv93}. These core hole states also impact the X-ray absorption spectrum (XAS) in Figure \ref{fig1}d, with additional influence from the 4$p$ density of states. The XAS final states are the intermediate states of the RIXS process.

\begin{figure}
\begin{center}
\includegraphics[width=3.4in]{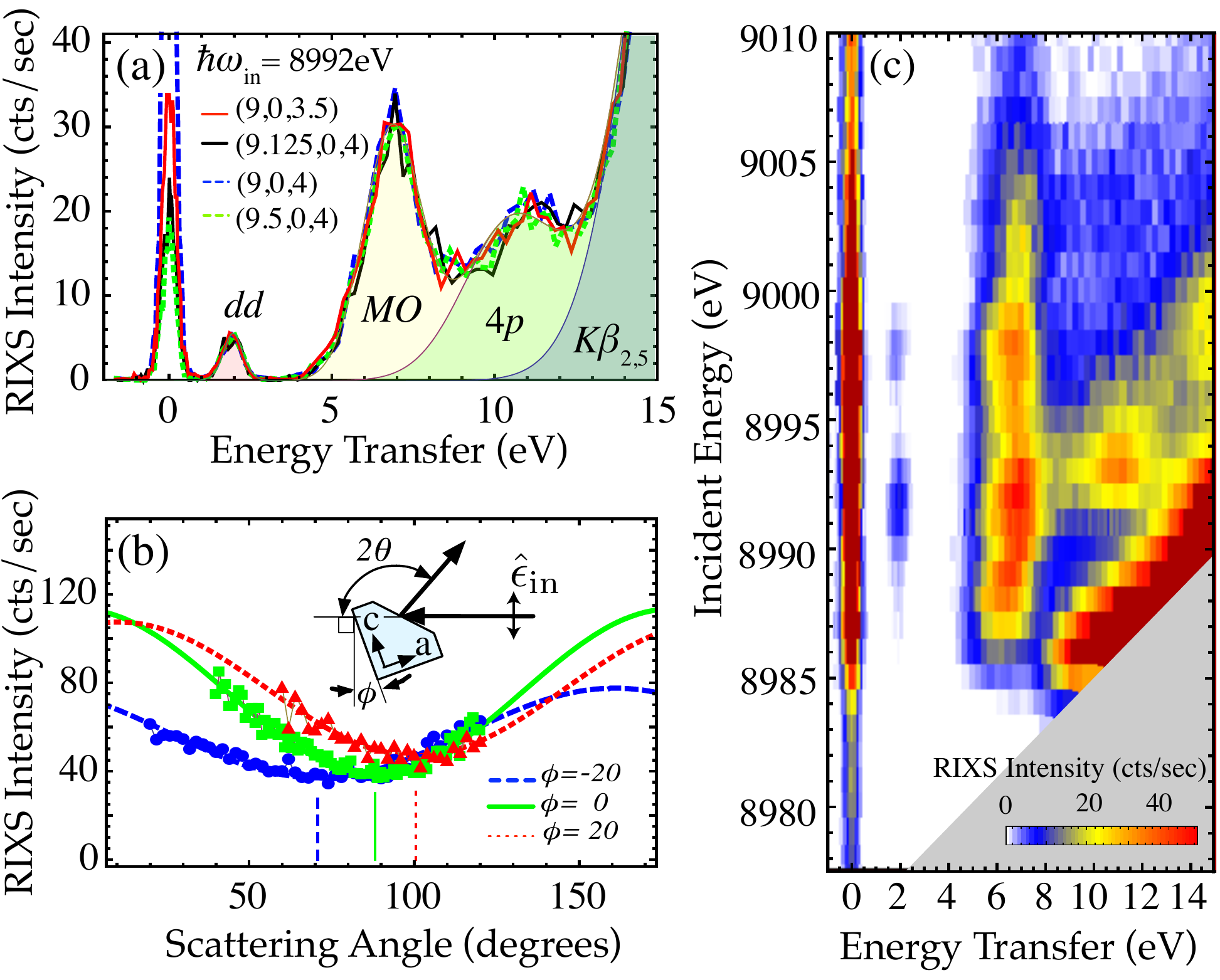}
\caption{(Color) (a) Momentum independence of the raw RIXS spectra for $\hbar\omega_\mathrm{in}$=8992 eV. Various spectrum components are indicated. (b) Angle dependence of the MO RIXS signal for three orientations ($\phi$) of the crystal ($\hbar\omega_\mathrm{in}$=8989 eV and $\Delta E$=6.75 eV). The lines indicate fits to the function $A$+$B \cos^2 (2\theta-\phi_0)$. (c) False color plot of RIXS spectrum with 2$\theta$=93$^\circ$, $\phi$=0, around {\bf{Q}}=(9,0,4). Data were corrected for self-absorption using a polarization-dependent mass absorption derived by merging the calculated value with the measured XAS spectra in the region of the edge. This step produces a spectrum which is closer to the RIXS cross section, but with systematic errors largest when the incident and/or final energy is near the edge at 8985 eV.}
\label{fig2}
\end{center}
\end{figure}

RIXS spectra were taken at the JAEA beamline BL11XU at SPring8, with incident polarization along the tetragonal $c$ axis, and energy resolution around 380 meV.
Figure \ref{fig2}a shows line scans of the RIXS intensity corresponding to several different momentum transfers, demonstrating the nondispersive (i.e., local) character of electronic excitations and corroborating our classification of \cbo\ as a quasi-zero-dimensional system. The unusual decoupling of momentum and angles enables us to isolate the intrinsic angle dependence of RIXS scattering in Fig. \ref{fig2}b, with angles defined in the inset. The intensity minimum follows the $c$-axis direction, suggesting an influence of the crystal orientation and polarization on signal level, perhaps due to 4$p$ electronic effects. A detailed understanding of these phenomena could aid future experimental work.

Figure \ref{fig2}c shows the incident-energy-dependent RIXS spectra. In contrast to the planar cuprates\cite{lu05,lu06}, only a small number of features appear as a function of transfered energy. The strongest feature appears as a diagonal line and corresponds to a constant scattered photon energy of $\sim$8976 eV. This fluorescence feature is the $K\beta_{2,5}$ emission line, which is understood to arise from processes including 3$d$$\rightarrow$$1s$ radiative transitions. The related peak at 10.9 eV is described with a calculation below.

\begin{figure}
\begin{center}
\includegraphics[width=2.8in]{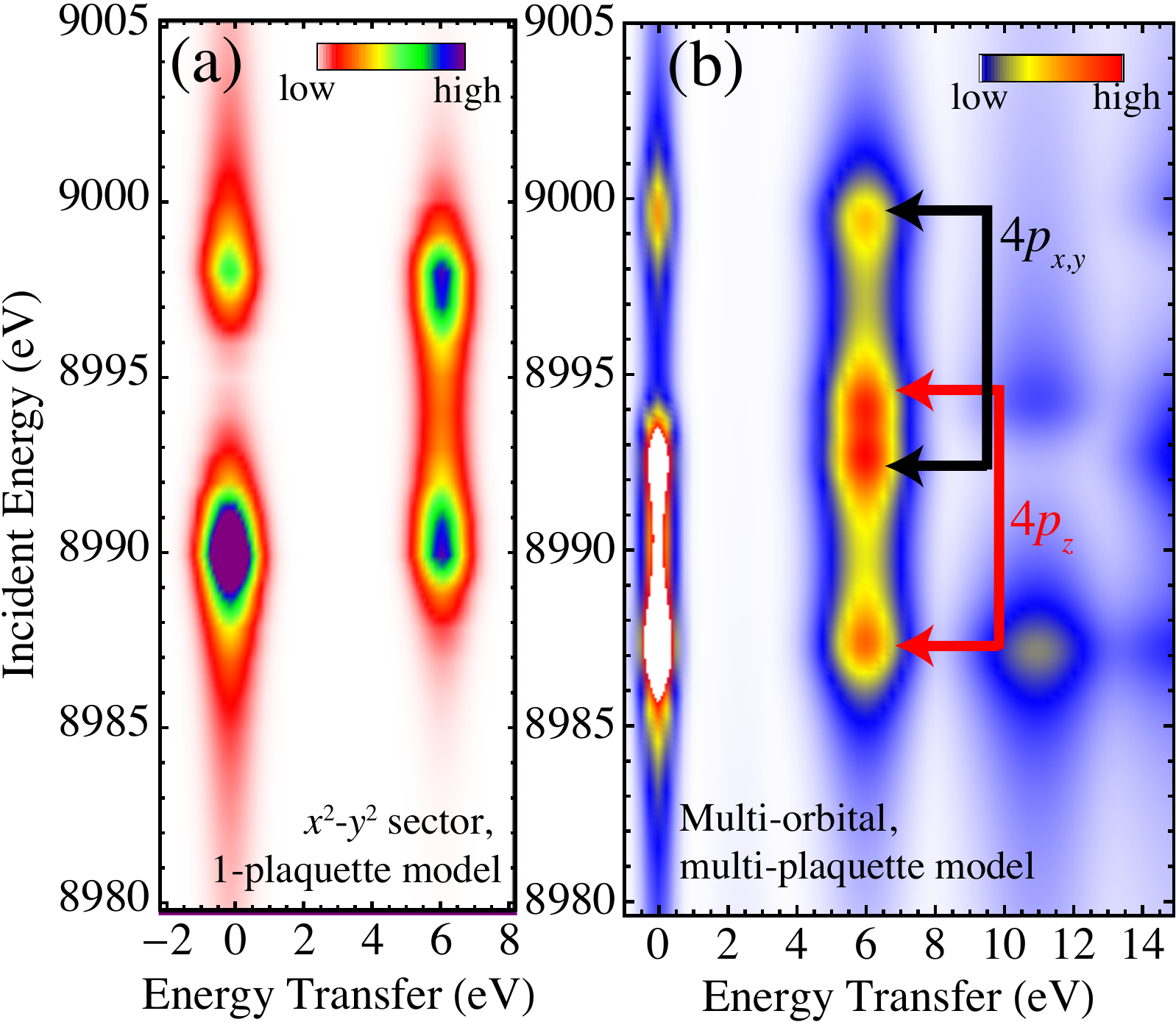}
\caption{(Color) (a) Calculated RIXS spectrum for the simple CuO$_4$ plaquette model. (b) Extended calculation of RIXS spectra, including all Cu 3$d$ and O 2$p$ orbitals, in the experimental configuration appropriate to one unit cell of \cbo\ (12 plaquettes).}
\label{fig3}
\end{center}
\end{figure}

Less understood prominent features having Raman-Stokes dispersion appear at about 1.9 and 6.8 eV energy transfer. We assign the spectral weight of the prominent $\sim$6.8 eV feature to the molecular orbital (MO) excitation, in accord with previous work \cite{kim04lico,yjk04mo}. This excitation forms a long streak in the incident-energy direction, a property that can be understood within a model including only two states of a CuO$_4$ plaquette \cite{vanv93}: the Cu 3$d_{x^2-y^2}$ ($|d^9\rangle$) and the $x^2$-$y^2$ combination of bonding O 2$p_\sigma$ orbitals surrounding the central Cu atom ($|d^{10}\underline{L}\rangle$). When hybridization is included in this model, the ground and excited states are formed as $|g\rangle$=$\cos{\alpha}|d^{10}\underline{L}\rangle$+$\sin{\alpha}|d^9\rangle$ and the orthogonal state $|MO\rangle$. In the presence of a Cu 1$s$ core hole, a new set of states $|WS\rangle$ and $|PS\rangle$ are formed. One can show the identity $\langle MO|WS\rangle\langle WS|g\rangle$=$-\langle MO|PS\rangle\langle PS|g\rangle$ is independent of the hopping or core hole interaction strength. This results in two equal-sized contributions to the inelastic signal which interfere constructively for incident photon energy $E_{WS}$$<$$\hbar\omega_\mathrm{in}$$<$$E_{PS}$, giving the appearance of the streak-like pattern in Figure \ref{fig3}a. The 4$p$ crystal field splitting produces two offset copies of this pattern, further elongating this streak, which is born out by our more advanced treatment below (Fig 3b). These properties are expected within conventional theoretical approaches.

In the energy transfer direction, the near-Gaussian line shape of the MO feature (Fig 2a) is unexpected for a long-lived excitation weakly coupled to its environment, and is unlikely to arise from inhomogeneous broadening in our narrow mosaic (0.002$^\circ$) crystals. One explanation is a Franck-Condon (F-C) effect, known to be present in C $K$ edge lineshapes\cite{hennies05}. Optical absorption spectra in \cbo\cite{pisarev04} (Figure 4a), show sharp ($\sim$1 meV) zero-phonon lines associated specifically with $dd$ excitations and many F-C sidebands. These arise when an electronic configurational change shifts the molecular potential before the nuclear coordinates can adjust, and show that at least between the ground and final states of optical absorption, electron-phonon coupling is quite strong in \cbo. The Gaussian line shape of the MO feature (Fig 2a) suggests that these F-C effects may play a role in transition metal $K$-edge RIXS, an effect that has not been directly observed to our knowledge. In a simple treatment, the deconvolved width $\Gamma_G$$\sim$0.76 eV would suggest that 2$\times$0.76/0.05$\sim$30 vibrational levels are excited at this electronic transition, assuming a typical phonon frequency $\hbar\omega_0$$\sim$50meV. This is in basic agreement with the findings of angle-resolved photoemission on high-$T_c$ parent compounds\cite{shen07}. F-C analysis of the MO feature in other systems could explain observed anomalies in the bond length dependence of the MO excitation energy peak across cuprate systems\cite{yjk04mo}.

The 1.9 eV feature resonates at 8992 and 8998 eV, has an energy-resolution-limited width, and is stronger than any similar feature reported previously \cite{kim04lico}. Higher-dimensional insulating systems (1D \cite{kim04lico}, 2D \cite{hasan00}, 3D \cite{doring04}) exhibit a feature assigned to a charge-transfer excitation near 2 eV. However, for \cbo\ such a nonlocal process should be absent due to the lack of electronically connected neighboring plaquettes. This excitation instead represents a local Cu $d$$d$ excitation. The unambiguous appearance of this type of excitation, which does not belong to the same representation as the ground state, is unexpected in a conventional treatment of $K$-edge RIXS, and therefore suggests the opening of a new scattering channel.

We have tested the possibility that trivial plaquette distortion and broken inversion symmetry can relax the selection rules of the RIXS process, rendering these transitions allowed. The Cu(2) plaquettes are distorted, but the Cu(1) sites are regular square planar units (exact $D_{4}$ symmetry). In a tight-binding picture of the 4$p$ band consistent with our modeling, the geometry in Figs. 2b and 4c ($\epsilon_i || 001$ and \textbf{Q}=(9,0,4)) would selectively highlight the undistorted sites at low incident energy ($\hbar\omega_\mathrm{in}$$<$8992 eV) and the distorted plaquettes at high incident energy ($\hbar\omega_\mathrm{in}$$>$8990 eV). Based on this information alone, one could argue that high incident energy peaks of the $dd$ resonance arise mainly from the Cu(2) plaquette distortion. Figures 4d-f compares the resonance profile of the $dd$ excitation with the same collected in a different geometry, (Fig 4c), where half of the distorted plaquettes resonate at low incident energy and the remaining plaquettes resonate at high incident energy. The negligible polarization dependence of the resonance profile on the scattering site implies that the $dd$ excitation is a robust, one-plaquette property, rather than resultant from Cu(2) distortions or inversion asymmetric crystal environments, which are very different for the two plaquette types.

We have performed an extended cluster calculation, now considering all orbitals of the valence system (O 2$p_\sigma$, 2$p_\pi$, 2$p_z$, and Cu $3d$ $e_g$ and $t_{2g}$ orbitals) and `core-spectator' system (Cu $1s$, 4$p_x$, 4$p_y$, and 4$p_z$ orbitals). We used a Hilbert space made of total spin +1/2 states with zero, one and two holes in the valence system, and respectively one, two and three electrons in the core-spectator system \footnote{Parameters used are: $(pd\sigma)$=1.54 eV, $(pd\pi)$=$(pp\sigma)$=1 eV, $(pp\pi)$=0.3 eV, $\epsilon_d$-$\epsilon_p$=4.0 eV, $Q$=7.0 eV.  The local tetragonal crystal field splitting $\epsilon_{4p_{x,y}}$-$\epsilon_{4p_z}$=5.56 eV, and $\epsilon_{4p_z}$=10.1 eV.}. With the numerical eigendata solutions of this system, we calculate the RIXS cross section of the 12 identical, decoupled plaquettes (Fig 1a) per unit cell using the Kramers-Heisenberg cross section:
\begin{equation}
\label{eqn:kh1}
\frac{d^2\sigma}{d\omega d\Omega} \propto \sum_{f,\nu} | \langle f| \mathcal{G}_\nu |g\rangle|^2 \delta(E_f-E_g-\hbar(\omega_{\rm{in}}-\omega_{\rm{sc}}))
\end{equation}
with the scattering operator for final photon polarization component $\nu$ $\in$ $\{ \pi,\sigma \}$ given as
\begin{equation*}
\label{ }
\mathcal{G}_{\nu}=\hat{\varepsilon}_{\rm{sc},\nu}\cdot \mathbf{J}^\dag(\mathbf{k}_{\rm{sc}}) \sum_i \frac{|i\rangle\langle i|}{\hbar \omega_{\rm{in}}-(E_i-E_g)-i\Gamma}\hat{\varepsilon}_{\rm{in}}\cdot \mathbf{J}(\mathbf{k}_{\rm{in}}).
\end{equation*}
$\hbar\omega_{\rm{in (sc)}}$, $\hat{\varepsilon}_{\rm{in (sc,\nu)}}$, and $\mathbf{k}_{\rm{in (sc)}}$ are the energy, polarization and momentum of the incident (scattered) photon, respectively. $\mathbf{J}(\mathbf{k})$=$\sum_{j} -i\nabla_{j} e^{i \mathbf{k}\cdot\mathbf{R}_{j}}$ is the momentum-dependent current operator, which we evaluate numerically for the differently oriented plaquettes using tabulated Hartree-Fock atomic wavefunctions\cite{rich63a}. The core-hole decay rate $\Gamma$ is set to 1 eV and the $\delta$ function in (\ref{eqn:kh1}) is replaced by a Gaussian of width $\gamma$=0.2+$0.06(E_f-Eg)$ eV.

Figure \ref{fig3}b shows the result of the calculation, which produces all observed features except the 1.9 eV excitation. The MO feature is robust, forming a long vertical streak, as expected. The extended basis additionally allows for features associated with 4$p$ electrons in the final state, similar to the 10.9 eV and $K\beta_{2,5}$ features. These features involve dipolar absorption and quadrupolar emission. With this identification, we estimate the ratio $R$=$I_{MO}/I_{K\beta_{2,5}}$, reflective of the dipole-dipole RIXS intensity relative to the dipole-quadrupole intensity, and find $R_{exp}$$<$$0.1$ and $R_{cluster}$$>$$6$. Such a large disparity does not come from errors in estimating the matrix elements, but can be understood in terms of dynamics beyond a one-plaquette cluster model. Because the 4$p$ photoelectron may hop off-site within a core-hole lifetime ($t_{4p}$$>$$\Gamma$), the likelihood for dipolar recombination is reduced, while the quadrupolar recombination channels are uneffected. This observed suppression of intensity points to a potentially strong material and pressure dependence of the RIXS scattering strength.

The appearance of the 1.9 eV feature requires further consideration beyond the simplest cluster model. 
Simple Cu 4$p_{x,y}$-O 2$p_\sigma$ hybridization effects are insufficient, but Coulomb interaction between the planar 4$p$ and 2$p_\sigma$ orbitals can break the selection rule of conventional RIXS in the intermediate state\cite{vernay08}, producing a $dd$ excitation. We have tested this in the cluster, and find that $U_{4p_{x,y}}$$_-$$_{2p_\sigma}$$\sim$1.5eV can produce approximately the correct relative strength between the $dd$ and $MO$ features, at the correct incident energy (Figure 4g). Alternatively, lattice vibrations can break symmetry\cite{hennies05}. While we cannot specify the precise cause of the 1.9 eV RIXS scattering in \cbo, we point out that the relative strength of the excitation at resonance ($I_{dd}/I_{MO}$$\sim$0.067) is similar in magnitude to the relative strength of the 2.2 eV feature in \lco\ ($I_{2.2eV}$/$I_{Total}$$\sim$0.08). This opens the possibility that the same symmetry-breaking process is at work in both cases.

Our measurements have allowed us to identify three distinct RIXS channels in \cbo, and we rank them here by relative strength: (i) the dipole-quadrupole channel responsible for the $K\beta_{2,5}$ and 10.9 eV features, (ii) the `core-only' channel responsible for the MO feature, and (iii) the unidentified channel responsible for the $dd$ excitations. 
Future theoretical and higher resolution experimental work could help identify the source of channel (iii) and address the symmetry of charge-transfer excitations in the high-$T_c$ superconducting systems.

\begin{figure}
\begin{center}
\includegraphics[width=3.4in]{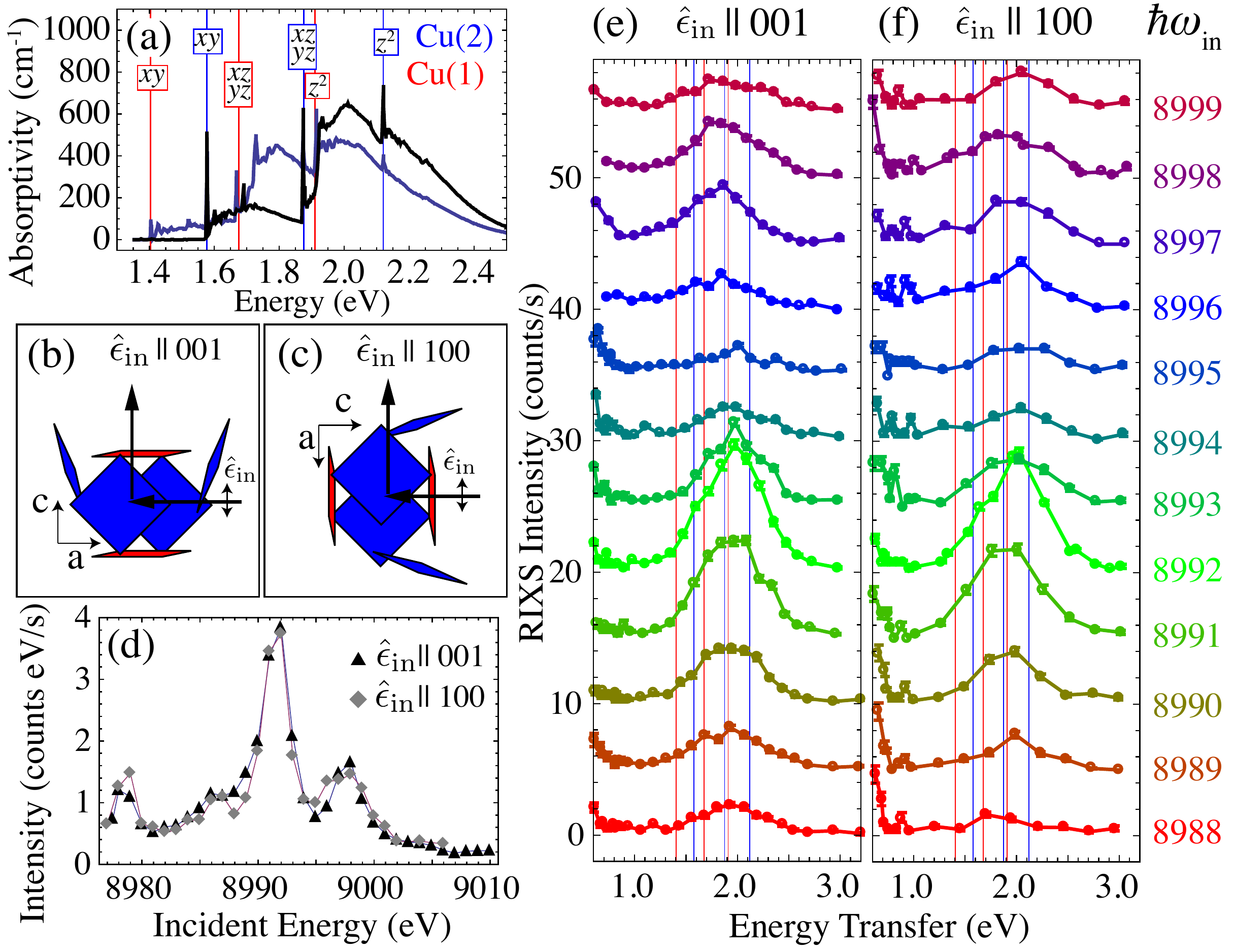}
\caption{(a) Optical absorption from \cite{pisarev04}, with assignments of the crystal field levels. (b,c) Cu(1) (red) and Cu(2) (blue) plaquette orientations in the two scattering geometries. (d) Resonance profile of the $dd$ excitation in the two geometries. (e,f), The $dd$ excitations energies (vertical lines) together with the low-energy feature seen in the RIXS data. (c) shown as vertical lines. Data in the $\epsilon_i ||$100 geometry were scaled to give the same $K\beta_{2,5}$ fluorescence intensity as in the $\hat{\epsilon}_{\mathrm{in}} ||$001 geometry.}
\label{ }
\end{center}
\end{figure}

We would like to acknowledge valuable conversations with J. P. Hill, S. Johnston, Y.-J. Kim, B. Moritz, B. S. Shastry, and F. Vernay.
This work was supported by the DOE under Contract No. DE-AC02-76SF00515 and by the NSF under Grant No. DMR-0705086.

\bibliography{rixs}
\end{document}